\documentclass[12pt]{iopart}
\expandafter\let\csname equation*\endcsname\relax
\expandafter\let\csname endequation*\endcsname\relax
\usepackage{graphicx}
\usepackage{siunitx}
\usepackage[version=4]{mhchem}
\usepackage{hyperref}
\usepackage[capitalize]{cleveref}
\usepackage[square,numbers]{natbib}
\usepackage{iopams}
\usepackage{caption}
\usepackage{subcaption}
\usepackage{multirow}
\usepackage{booktabs}
\usepackage{algorithm}
\usepackage{algpseudocode}
\usepackage[title]{appendix}

\begin{document}

\title[Agentic Global Optimization of Site-Disordered Crystalline Materials]{Agentic Global Optimization of Site-Disordered Crystalline Materials}

\author{Martin Hoffmann Petersen$^{1,2}$*, Zekun Ren$^2$, and Kedar Hippalgaonkar$^{1,2}$*}

\address{1 School of Materials Science and Engineering, Nanyang Technological University, Singapore 639798}
\address{2 Berkeley Education Alliance for Research in Singapore, CREATE Tower, Singapore 138602}

\ead{martin.petersen@ntu.edu.sg, kedar@ntu.edu.sg}

\begin{abstract}
Site-disordered crystals, in which several elements partially occupy the same atomic site, cannot be simulated atomistically without first resolving an explicit ordering, and the number of possible orderings grows combinatorially with supercell size. Because each candidate requires an expensive structural relaxation, locating the global-minimum ordering is a costly combinatorial search. Here, we present an agentic search in which a surrogate model, fitted to previously evaluated configurations, selects each subsequent candidate by balancing exploitation, uncertainty, and novelty. Four surrogates are implemented within a shared framework: an ExtraTrees ensemble, Bayesian ridge regression, sparse ARD regression, and an ALiEGNN-inspired graph ensemble. These are benchmarked against three traditional methods: random walk, Markov chain Monte Carlo, and a genetic algorithm across 14 test cases spanning 12 crystal structures, with exhaustively enumerated global minima as ground truth. The agentic surrogates reach the global minimum in 94\% of runs, compared with 55\% for the traditional methods, with a median 4.3-fold reduction in evaluations relative to the best traditional method. Bayesian ridge regression succeeds in all runs with the fewest evaluations. Applied to \ce{Na_{2/3}FePO4}, with coupled Na/vacancy configurational disorder and \ce{Fe^{2+}}/\ce{Fe^{3+}} charge disorder, the framework recovers the experimentally reported ordering when charge states are encoded in the surrogate representation, demonstrating its extension to multi-sublattice disorder.
\end{abstract}

\section{Introduction}
A crystal is characterized by a periodic arrangement of atoms in space, defined by its unit cell and governed by symmetry operations derived from the crystal's space group and site symmetries. However, this ideal symmetry often breaks down with the introduction of doping, an essential strategy used to enhance the performance of functional materials\cite{Zhou2025,wang2024disordered}. In such cases, atomic sites may be partially occupied by different elements, introducing atomic site-disorder that prevents a straightforward description of the crystal structure with a single unit cell\cite{Souvignier2016, Antypov2025,SWORD}. Similarly, during experimental synthesis, even nominally ordered materials can exhibit partial occupancies as atoms redistribute among several symmetry-equivalent sites\cite{Ndione2013,Haiwen_paper,Leeman2024}. All of this contributes to the fact that nearly 50\% of all experimentally verified crystal structures in the Inorganic Crystal Structure Database (ICSD)\cite{icsd} exhibit some degree of atomic site-disorder\cite{Antypov2025}.

The biggest issue with site-disordered crystal structures, from an atomistic point of view, is the inaccessibility of performing atomistic simulations using either first-principles calculations or Machine Learning Interatomic Potentials (MLIPs), since the partial occupancy of elements makes the electronic configuration at each of the Wyckoff sites difficult to define\cite{Yang2016,yong2024dismai,Peng2026}. The partial occupancy of the elements can create a probabilistic issue, where different orderings of elements result in approximately the same energy, resulting in a solid solution, while it can also be an optimization issue, as there might be one ordering that is more energetically favorable than others\cite{VIRP,Dis_gen}. Critically, these two regimes are not merely a formal distinction. The physical properties of a material, such as its formation energy, ionic conductivity, or magnetic ordering, can depend sensitively on which specific configuration is realized\cite{zhong2024effect,Zhou2006,Wang2012}. Identifying whether a system settles into one energetically preferred ordering or a near-degenerate solid solution is therefore a prerequisite for predicting how the material will actually behave.

Traditionally, approaches such as cluster expansion (CE)\cite{chang2019clease,de1994cluster} and special quasirandom structures (SQS)\cite{sqs} have been employed to study site disorder using first-principles calculations. Both remain valuable tools, particularly for capturing phase transitions and finite-temperature behavior, where CE's fitted energy model enables efficient sampling across many configurations and temperatures. However, neither was designed as a general search procedure, and locating the single globally optimal ordering can be achieved more efficiently through direct optimization over configuration space\cite{Elsborg2024}.

This motivates treating site-disorder resolution directly as a global optimization problem: given a fixed composition and a lattice, search the configurational space for the ordering that minimizes total energy, without initially assuming a functional form for how that energy depends on configuration. This problem is not new, and methods such as Markov chain Monte Carlo (MCMC)\cite{Metropolis1953,Ismail2024} and genetic algorithms (GA)\cite{Lysgaard2013,Jennings2019} have been utilized to find globally optimized structures using either first-principles calculations or surrogates for energy or property prediction. Recently, foundational MLIPs have changed the paradigm for evaluating the search space, as they can now perform geometry optimization faster than first-principles calculations while keeping accuracy close to that of first-principles calculations\cite{m3gnet,Mace,yang2024mattersim}. That being the case, when considering larger crystal structures, which are required for some disordered systems, even MLIPs can become a computational bottleneck. Clever global optimization algorithms are therefore needed to minimize the number of evaluation steps.

Agentic artificial intelligence (AI), in which an autonomous loop plans, acts, and updates its strategy from observed outcomes, offers a natural framework for sample-efficient search\cite{PUDA,el_agente}. Instead of drawing candidates from a fixed sampling policy, an agentic search fits a surrogate model to the configurations evaluated so far. It then selects each next candidate by balancing three factors: exploitation of the current best estimate, the uncertainty of that estimate, and novelty relative to configurations already explored. This adaptive, closed-loop strategy mirrors the broader development of autonomous and self-driving approaches in materials discovery\cite{TetsassiFeugmo2026,bolle2020autonomous,ai_roadmap}.

\begin{figure}[t]
\centering
\includegraphics[width=0.95\textwidth]{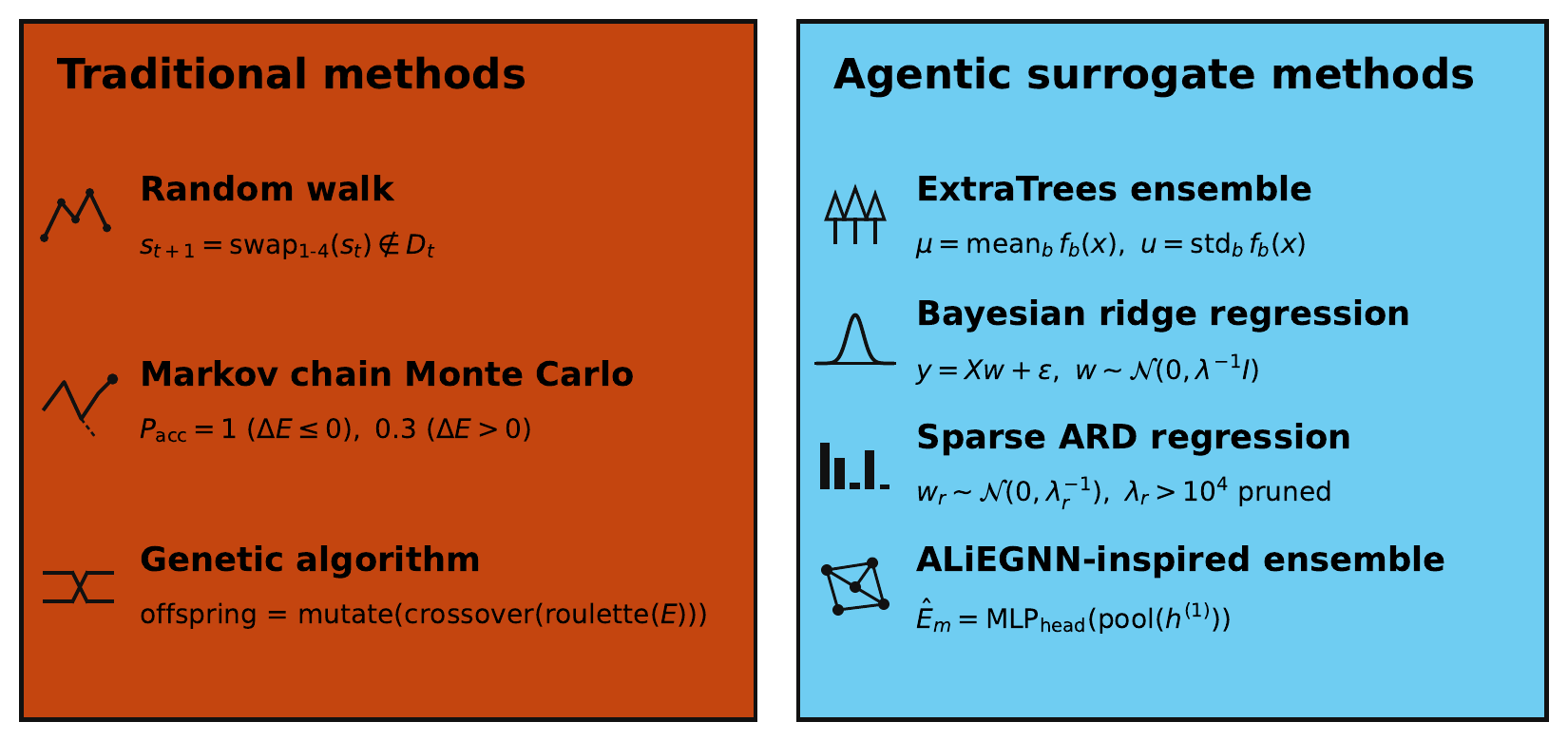}
\caption{Overview of the seven global optimization strategies compared in this work. Left: the three traditional baselines (random walk, Markov chain Monte Carlo, and a genetic algorithm), each shown with its proposal or acceptance rule. Right: the four agentic surrogate methods (ExtraTrees ensemble, Bayesian ridge regression, sparse ARD regression, and an ALiEGNN-inspired graph ensemble), each shown with its defining model expression.}
\label{fig:methods}
\end{figure}

In this work, we investigate whether an agentic search can locate the global-minimum ordering of site-disordered crystal structures in fewer evaluations than classical global optimization strategies. As illustrated in \cref{fig:methods}, we implement four surrogate models within a shared agentic framework: (i) an ExtraTrees ensemble, (ii) Bayesian ridge regression, (iii) sparse automatic relevance determination (ARD) regression\cite{pedregosa2011scikit}, and (iv) an ensemble of compact ALiEGNN-inspired graph neural networks\cite{klat_surrogate}. These are benchmarked against random walk, MCMC, and GA across 14 test cases spanning 12 distinct crystal structures. For each case, all symmetry-inequivalent configurations were enumerated exhaustively, which establishes the global-minimum structure as ground truth\cite{chang2019clease,ASE,Na_dataset}. All methods share the same energy evaluator (TensorNet-MATPES-PBE-v2)\cite{Matgl,matpes} and a maximum budget of 200 energy evaluations per system, and the agentic searches begin from a single random initial structure, followed by initial selection of eight configurations. As global optimization algorithms are inherently stochastic, each method is repeated over 5 independent runs. Performance is assessed by whether the true global minimum is recovered and how quickly the search converges over these runs.

The best agentic surrogate reduces the number of evaluations by a median factor of 4.4 relative to the best traditional method. To test the approach beyond the benchmark, we apply the Bayesian ridge search to \ce{Na_{2/3}FePO4}, in which Na/vacancy configurational disorder and \ce{Fe^{2+}}/\ce{Fe^{3+}} charge disorder coexist. Combining a charge-state-aware MLIP \cite{charge_ordering} with a charge-encoded surrogate representation, the agentic search recovers the experimentally observed ordering \cite{kyoto,kyoto_cif}. These results establish agentic surrogate-guided search as an efficient alternative to traditional global optimization. Its adaptive nature makes it well suited to complex systems with multiple coupled disordered sublattices, which are increasingly costly to resolve with traditional methods.

\section{Results}
\label{sec:results}

\subsection{Global optimization benchmark}

\begin{figure}[b]
    \centering
    \includegraphics[width=0.95\textwidth]{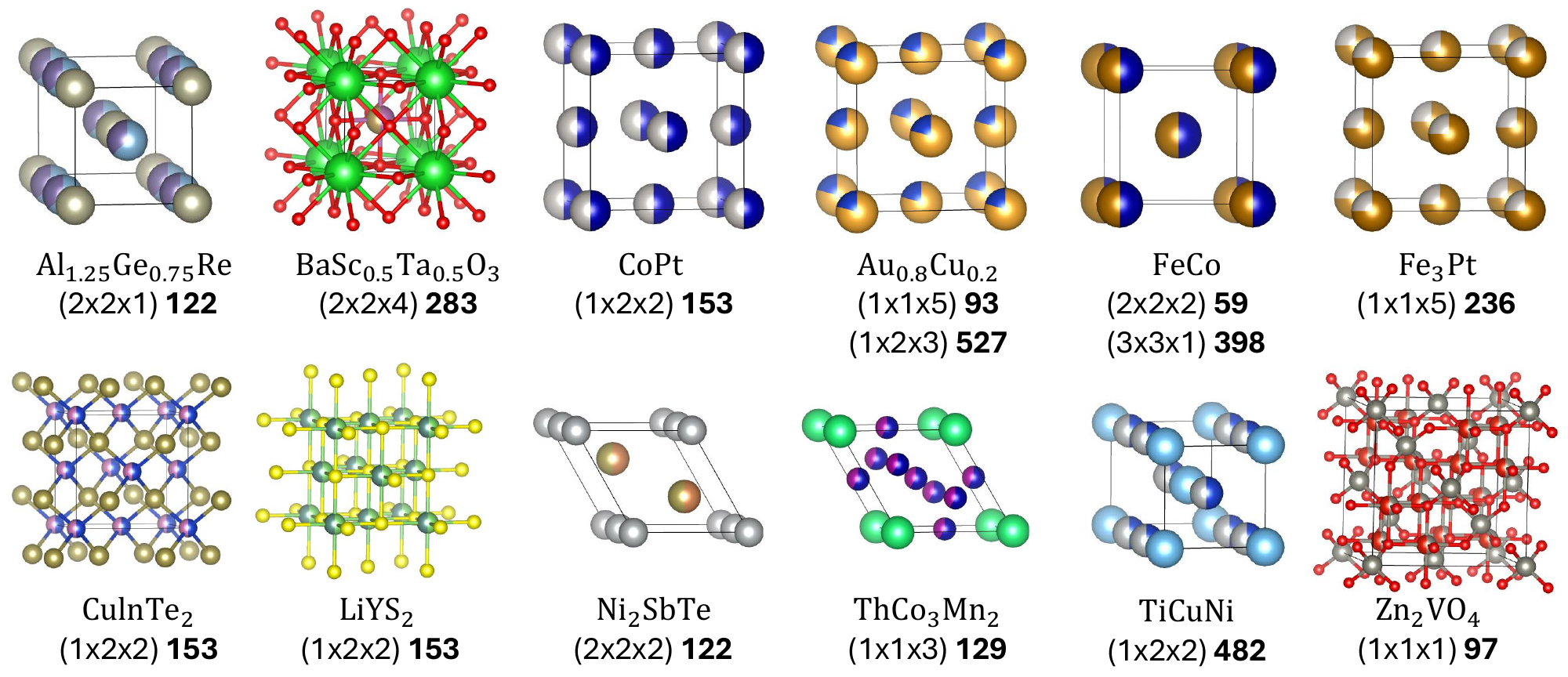}
    \caption{The twelve site-disordered crystal systems used in the global optimization benchmark. The supercell size is given beneath each formula, and the bold number is the number of symmetry-inequivalent configurations enumerated for that supercell.}
    \label{fig:crystals}
\end{figure}

The twelve systems shown in \cref{fig:crystals} span a broad range of chemistries, coordination environments, and configuration-space sizes, from 59 symmetry-inequivalent configurations (\ce{FeCo}, $2\times2\times2$) to 527 (\ce{Au_{0.8}Cu_{0.2}}, $1\times2\times3$). For \ce{FeCo} and \ce{Au_{0.8}Cu_{0.2}}, two supercell sizes are considered to test whether relative method performance depends on the size of the enumerated space rather than on chemistry alone, giving fourteen test cases in total.

\begin{figure}[h]
    \centering
    \includegraphics[width=0.75\textwidth]{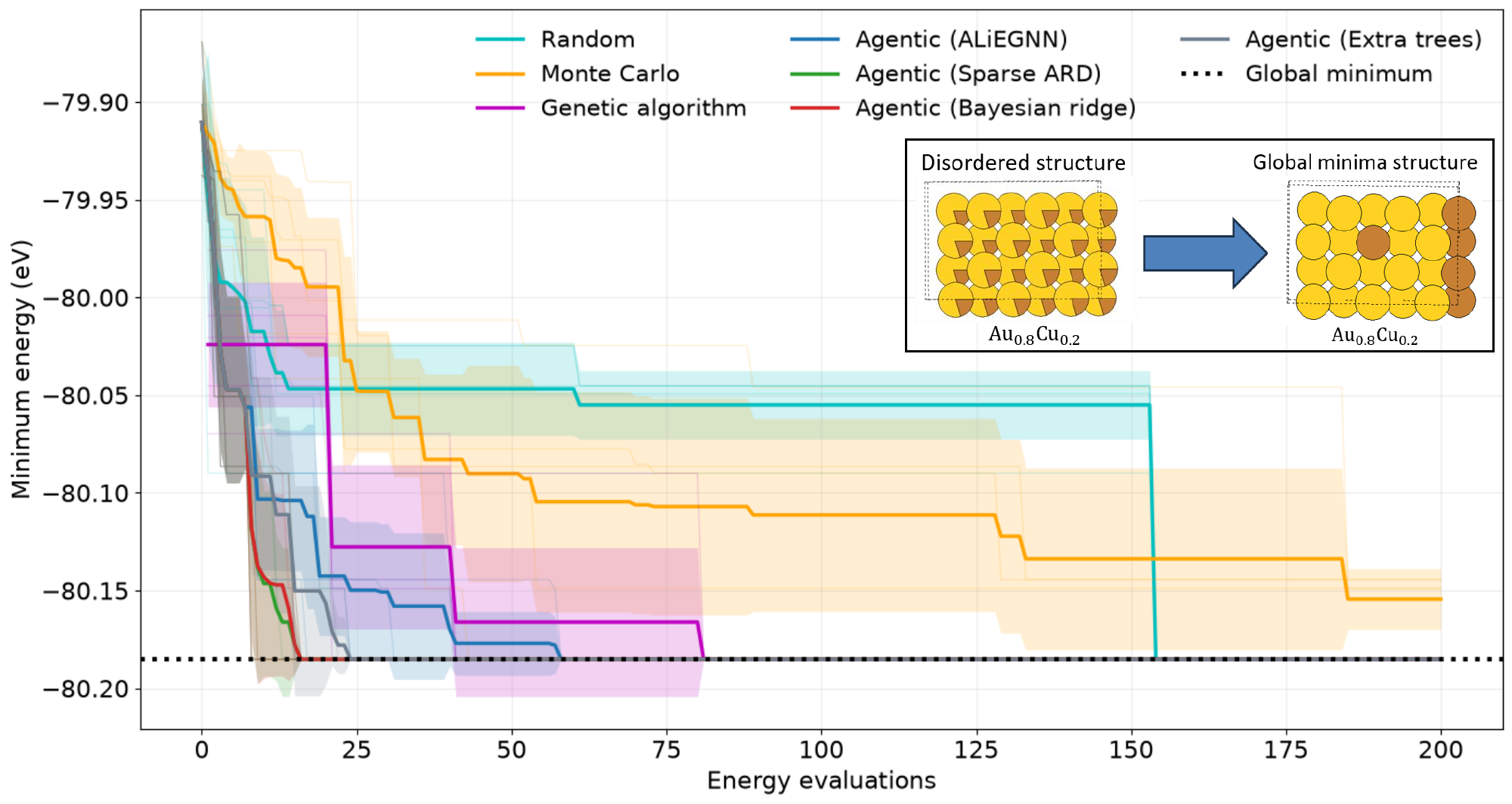}
    \caption{Best-so-far minimum energy as a function of the number of energy evaluations for all seven methods on \ce{Au_{0.8}Cu_{0.2}} ($1\times2\times3$), averaged over five independent runs (shaded bands: $\pm1\sigma$). The dotted line marks the enumerated global minimum. Inset: the disordered starting configuration and the recovered global-minimum ordering.}
    \label{fig:example}
\end{figure}

\ce{Au_{0.8}Cu_{0.2}} ($1\times2\times3$) illustrates the separation between the two method classes particularly clearly (\cref{fig:example}). The three descriptor-based surrogates (ARD, Bayesian ridge, and ExtraTrees) reach the enumerated global minimum in every run, requiring on average only 12-18 evaluations, i.e. shortly after the eight-structure warm-up. The ALiEGNN ensemble also succeeds in every run but requires $35.2 \pm 14.3$ evaluations. The GA reaches the minimum in all runs after $69.0 \pm 50.0$ evaluations, passing through clearly suboptimal plateaus, consistent with its methodology that a population must first be discovered and then it propagates towards favourable configurations. Random walk and MCMC perform worst, as random walk succeeds in three of five runs, after 155 evaluations, and only one of five MCMC runs reaches the global minimum. All seven methods recover the same global-minimum structure, or a symmetry-equivalent one, in at least one run, regardless of their search strategy. This agreement confirms both that the exhaustive enumeration provides a well-defined ground truth for the benchmark, and that global optimization strategies can locate this minimum without enumerating the full configuration space. The methods therefore differ in search efficiency, not in the structure they ultimately identify.

\begin{figure}[h]
    \centering
    \includegraphics[width=0.95\textwidth]{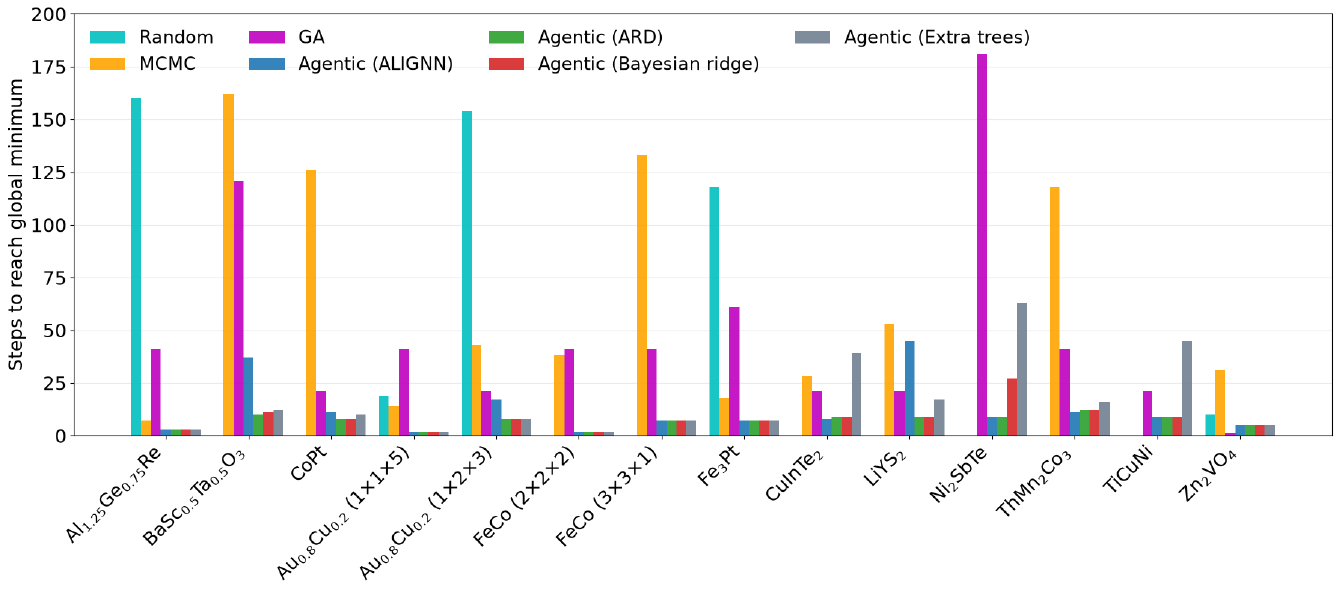}
    \caption{Number of energy evaluations required by the fastest of five independent runs of each method to first reach the global-minimum configuration, shown for all fourteen benchmark cases. Missing bars indicate that no run of that method reached the global-minimum structure within the 200-evaluation budget.}
    \label{fig:histogram}
\end{figure}

Across all fourteen benchmark cases, summarized in \cref{fig:histogram} and shown in full in \cref{app:all_runs}, the four agentic surrogates consistently require fewer energy evaluations to reach the global minimum than the three traditional baselines. Random walk and MCMC are the least efficient, frequently requiring more than 100 evaluations and in several cases failing to converge within the 200-evaluation budget. The GA is the strongest traditional method, but its performance depends strongly on the crystal system: even its fastest run ranges from 181 evaluations on \ce{Ni2SbTe2}, where only one of five runs reaches the global minimum, down to a single evaluation on \ce{Zn2VO4}, where the global minimum was already present in the initial population. The ExtraTrees and ALiEGNN ensembles are markedly faster than the traditional baselines but show greater variability between systems, with fastest runs requiring up to 64 evaluations on \ce{Ni2SbTe2} for ExtraTrees and 46 on \ce{LiYS2} for ALiEGNN. In contrast, the fastest runs of Bayesian ridge and sparse ARD regression remain below 25 evaluations in every system, with the single exception of Bayesian ridge on \ce{Ni2SbTe2} (28 evaluations). Overall, the efficiency gains of agentic search hold consistently across chemically and structurally diverse site-disordered systems, rather than being confined to a few favourable cases.

\Cref{tab:stats} summarizes the results for each system, with the final row averaging the per-system means so that every system contributes equally. Per-run evaluation counts for every method are given in \cref{app:seeds}. The agentic surrogates reach the global minimum in 264 of 280 runs (94.3\%), whereas the traditional baselines succeed in only 116 of 210 runs (55.2\%). The agentic surrogates are also consistently faster, as every agentic surrogate requires fewer evaluations on average (19.5--36.8) than every traditional baseline (78.2--115.8). Bayesian ridge regression stands out as the most effective surrogate overall. It is the only method to reach the global minimum in all 70 runs, and it does so with the fewest evaluations on average ($19.5 \pm 11.1$), roughly 4.0 times fewer than the GA, the strongest traditional baseline ($78.2 \pm 40.2$). Bayesian ridge regression spread across systems is also the smallest of all methods, making it the most consistent method overall, and it requires fewer evaluations than the GA in every one of the 14 systems, by a median factor of 4.3 (range 1.8--11.9). Although ALiEGNN and ExtraTrees outperform the other methods in a few cases (\ce{ThCo3Mn2} for ALiEGNN, and \ce{Fe3Pt} and \ce{Zn2VO4} for ExtraTrees), they require almost twice as many evaluations as Bayesian ridge on average (ALiEGNN: $36.8 \pm 21.3$; ExtraTrees: $32.8 \pm 26.9$) and show a larger spread, making them less efficient overall. Sparse ARD regression is, however, the fastest method in 8 of the 14 systems, including \ce{Al_{1.25}Ge_{0.75}Re}, where it requires only $9.4 \pm 2.9$ evaluations. In these systems, its advantage over Bayesian ridge is small (typically fewer than 2 evaluations), whereas in others, such as \ce{Au_{0.8}Cu_{0.2}} ($1\times1\times5$), \ce{Fe3Pt}, and \ce{Zn2VO4}, ARD is considerably slower. This is consistent with its sparse relevance selection, which can be highly efficient when the correct cluster correlations are identified early but less robust when they are not. Bayesian ridge regression, by contrast, is rarely the fastest method but never far from it, and this consistency makes it the better overall choice as it is the most reliable surrogate across chemically diverse systems and therefore the recommended default.

To test whether these differences are statistically significant, we first compare the recommended default, Bayesian ridge, directly with the GA, the strongest traditional baseline. A one-sided paired Wilcoxon signed-rank test \cite{Wilcoxon1945} over the 14 systems confirms that Bayesian ridge requires significantly fewer evaluations ($p = 6.1\times10^{-5}$, $n = 14$), being faster in every system, from a factor of 1.8 on \ce{ThCo3Mn2} (44.8 versus 81.0 evaluations) to 11.9 on \ce{FeCo} ($2\times2\times2$; 5.8 versus 69.0). A similar result holds at the level of method classes. For each system, we compare the fastest agentic surrogate with the fastest traditional baseline. The fastest agentic method is sparse ARD in eight systems, Bayesian ridge in three, ExtraTrees in two, and the ALiEGNN ensemble in one, while the fastest traditional method is the GA in nine systems, MCMC in four, and random walk in one. In this comparison, the agentic methods are faster by a median factor of 4.84 (mean 4.82; $p = 6.1\times10^{-5}$, $n = 14$), ranging from 1.37 on \ce{Au_{0.8}Cu_{0.2}} ($1\times1\times5$), where Bayesian ridge requires 14.6 evaluations against 20.0 for random walk, to 9.68 on \ce{BaSc_{0.5}Ta_{0.5}O3}, where sparse ARD requires 15.6 evaluations against 151.0 for the GA. No traditional method outperforms the best agentic surrogate in any system.

\begin{table}[h]
\centering
\caption{Energy evaluations required to first reach the enumerated global minimum for each method and benchmark system, given as mean $\pm$ standard deviation over successful runs, with the number of successful runs out of five in parentheses. A dash indicates that no run reached the global minimum within the evaluation budget. The lowest mean for each system is shown in bold. The final row gives the mean $\pm$ standard deviation of the per-system means, taken over all systems in which the method succeeded at least once, together with the total number of successful runs out of 70.}
\label{tab:stats}
\resizebox{\textwidth}{!}{%
\begin{tabular}{@{}lccc|cccc@{}}
\toprule
 & \multicolumn{3}{c|}{Traditional} & \multicolumn{4}{c}{Agentic} \\
System & Random walk & MCMC & GA & Bayesian ridge & Sparse ARD & ALiEGNN & ExtraTrees \\
\midrule
\ce{Al_{1.25}Ge_{0.75}Re} & 188.2 $\pm$ 13.6 (5/5) & 51.6 $\pm$ 38.3 (5/5) & 65.0 $\pm$ 23.3 (5/5) & 11.2 $\pm$ 5.9 (5/5) & \textbf{9.4 $\pm$ 2.9 (5/5)} & 41.4 $\pm$ 34.9 (5/5) & 14.0 $\pm$ 6.1 (5/5) \\
\ce{BaSc_{0.5}Ta_{0.5}O3} & \multicolumn{1}{c}{--- (0/5)} & 163.0 $\pm$ 0.0 (1/5) & 151.0 $\pm$ 30.0 (2/5) & 17.4 $\pm$ 4.3 (5/5) & \textbf{15.6 $\pm$ 5.4 (5/5)} & 53.0 $\pm$ 11.0 (5/5) & 17.2 $\pm$ 2.9 (4/5) \\
\ce{CoPt} & \multicolumn{1}{c}{--- (0/5)} & 140.0 $\pm$ 11.0 (3/5) & 33.0 $\pm$ 9.8 (5/5) & 12.4 $\pm$ 2.1 (5/5) & \textbf{11.8 $\pm$ 1.9 (5/5)} & 21.6 $\pm$ 8.1 (5/5) & 22.6 $\pm$ 8.4 (5/5) \\
\ce{Au_{0.8}Cu_{0.2}} ($1{\times}1{\times}5$) & 20.0 $\pm$ 0.0 (4/5) & 40.8 $\pm$ 35.0 (4/5) & 85.0 $\pm$ 38.8 (5/5) & \textbf{14.6 $\pm$ 7.3 (5/5)} & 35.0 $\pm$ 31.4 (5/5) & 28.2 $\pm$ 17.4 (5/5) & 23.0 $\pm$ 13.1 (5/5) \\
\ce{Au_{0.8}Cu_{0.2}} ($1{\times}2{\times}3$) & 155.0 $\pm$ 0.0 (5/5) & 44.0 $\pm$ 0.0 (1/5) & 45.0 $\pm$ 19.6 (5/5) & 12.2 $\pm$ 3.5 (5/5) & \textbf{11.8 $\pm$ 3.2 (5/5)} & 35.2 $\pm$ 14.3 (5/5) & 17.8 $\pm$ 5.7 (5/5) \\
\ce{FeCo} ($2{\times}2{\times}2$) & \multicolumn{1}{c}{--- (0/5)} & 53.5 $\pm$ 14.5 (2/5) & 69.0 $\pm$ 16.0 (5/5) & 5.8 $\pm$ 3.0 (5/5) & \textbf{5.6 $\pm$ 2.7 (5/5)} & 5.8 $\pm$ 3.0 (5/5) & 6.6 $\pm$ 4.5 (5/5) \\
\ce{FeCo} ($3{\times}3{\times}1$) & \multicolumn{1}{c}{--- (0/5)} & 145.0 $\pm$ 11.0 (2/5) & 57.0 $\pm$ 8.0 (5/5) & 10.6 $\pm$ 1.4 (5/5) & \textbf{10.4 $\pm$ 1.4 (5/5)} & 30.8 $\pm$ 17.3 (5/5) & 14.8 $\pm$ 5.3 (5/5) \\
\ce{Fe3Pt} & 119.0 $\pm$ 0.0 (5/5) & 111.0 $\pm$ 66.1 (5/5) & 87.7 $\pm$ 18.9 (3/5) & 27.8 $\pm$ 28.3 (5/5) & 58.4 $\pm$ 69.7 (5/5) & 53.2 $\pm$ 60.3 (5/5) & \textbf{15.0 $\pm$ 7.0 (2/5)} \\
\ce{CuInTe2} & \multicolumn{1}{c}{--- (0/5)} & 63.8 $\pm$ 33.8 (4/5) & 29.0 $\pm$ 16.0 (5/5) & \textbf{12.2 $\pm$ 2.9 (5/5)} & 12.4 $\pm$ 2.1 (5/5) & \textbf{12.2 $\pm$ 3.1 (5/5)} & 55.5 $\pm$ 15.5 (2/5) \\
\ce{LiYS2} & \multicolumn{1}{c}{--- (0/5)} & 54.0 $\pm$ 0.0 (1/5) & 61.0 $\pm$ 35.8 (5/5) & \textbf{12.4 $\pm$ 2.9 (5/5)} & 18.4 $\pm$ 6.1 (5/5) & 67.0 $\pm$ 21.0 (2/5) & 20.0 $\pm$ 2.0 (2/5) \\
\ce{Ni2SbTe2} & \multicolumn{1}{c}{--- (0/5)} & \multicolumn{1}{c}{--- (0/5)} & 181.0 $\pm$ 0.0 (1/5) & 37.6 $\pm$ 7.2 (5/5) & \textbf{34.0 $\pm$ 23.7 (5/5)} & 84.2 $\pm$ 77.1 (4/5) & 83.2 $\pm$ 19.1 (5/5) \\
\ce{ThCo3Mn2} & \multicolumn{1}{c}{--- (0/5)} & 119.0 $\pm$ 0.0 (1/5) & 81.0 $\pm$ 32.7 (3/5) & 44.8 $\pm$ 37.8 (5/5) & 44.0 $\pm$ 29.1 (4/5) & \textbf{25.0 $\pm$ 14.0 (5/5)} & 71.2 $\pm$ 49.9 (4/5) \\
\ce{TiCuNi} & \multicolumn{1}{c}{--- (0/5)} & \multicolumn{1}{c}{--- (0/5)} & 73.0 $\pm$ 53.1 (5/5) & 29.2 $\pm$ 15.8 (5/5) & \textbf{28.0 $\pm$ 14.7 (5/5)} & 44.0 $\pm$ 39.4 (5/5) & 84.8 $\pm$ 21.0 (5/5) \\
\ce{Zn2VO4} & 96.8 $\pm$ 85.3 (4/5) & 80.0 $\pm$ 37.0 (5/5) & 77.0 $\pm$ 63.7 (5/5) & 25.0 $\pm$ 22.9 (5/5) & 60.8 $\pm$ 59.8 (5/5) & 14.2 $\pm$ 4.3 (5/5) & \textbf{13.6 $\pm$ 6.5 (5/5)} \\
\midrule
Mean over systems & 115.8 $\pm$ 57.1 (23/70) & 88.8 $\pm$ 42.4 (34/70) & 78.2 $\pm$ 40.2 (59/70) & \textbf{19.5 $\pm$ 11.1 (70/70)} & 25.4 $\pm$ 17.8 (69/70) & 36.8 $\pm$ 21.3 (66/70) & 32.8 $\pm$ 26.9 (59/70) \\
\bottomrule
\end{tabular}%
}
\end{table}

\subsection{Multi-sublattice disorder: \ce{Na_{2/3}FePO4}}

\begin{figure}[h]
\centering
\includegraphics[width=0.8\textwidth]{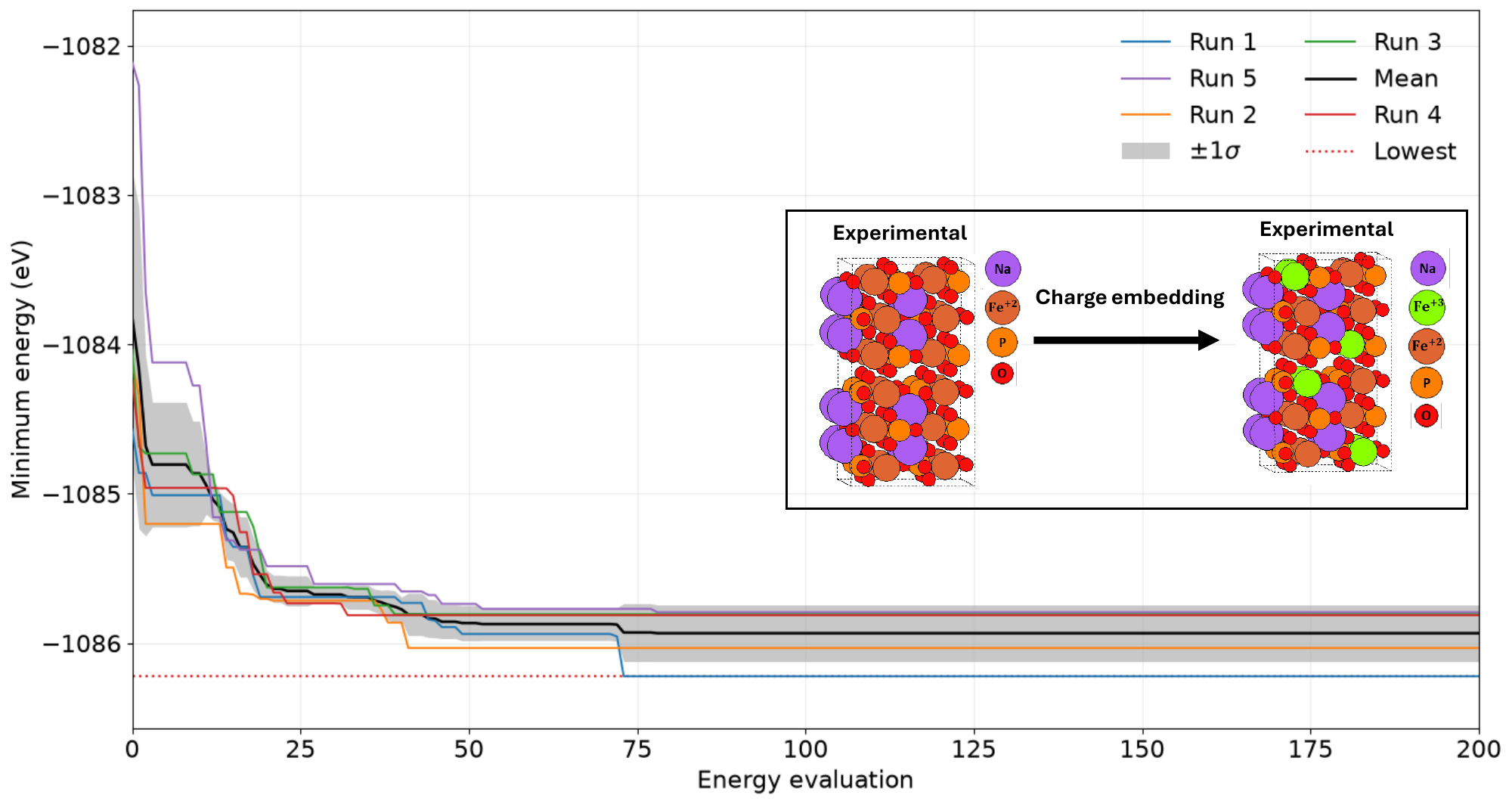}
\caption{Best-so-far minimum energy for five independent Bayesian-ridge search runs on \ce{Na_{2/3}FePO4}, searching jointly over Na/vacancy ordering and \ce{Fe^{2+}}/\ce{Fe^{3+}} charge ordering. The dotted line marks the energy of the experimentally reported structure. Inset: the experimental structure before (left) and after (right) embedding the \ce{Fe^{2+}}/\ce{Fe^{3+}} charge states.}
\label{fig:nafepo4}
\end{figure}

\ce{Na_{2/3}FePO4} is considerably more demanding than the benchmark systems because it contains two coupled sources of disorder: (i) Na/vacancy configurational ordering and (ii) \ce{Fe^{2+}}/\ce{Fe^{3+}} charge ordering, which arises as Fe changes oxidation state to compensate for \ce{Na+} removal \cite{Polyanion_dataset}. The two are linked by charge neutrality, since each Na vacancy is compensated by one \ce{Fe^{3+}}, so the numbers of vacancies and \ce{Fe^{3+}} ions, and likewise of \ce{Na+} and \ce{Fe^{2+}} ions, are fixed relative to each other. The search must therefore handle not only two kinds of disorder but also the correlation between them. Because charge ordering alone can produce large energy differences \cite{charge_ordering,Zhou2006,Malica2025}, a representation that tracks only Na/vacancy occupancy cannot distinguish orderings that differ solely in charge assignment. We therefore search over both sublattices jointly, embedding the \ce{Fe^{2+}}/\ce{Fe^{3+}} assignment directly into the surrogate representation alongside the Na/vacancy occupancy as well as performing the energy evaluation using a charge-encoded MLIP trained on Fe-based polyanion compounds \cite{charge_ordering}.

Bayesian ridge regression, the strongest agentic surrogate in the benchmark, is used for this system, with five independent runs. As shown in \cref{fig:nafepo4}, one of the five runs recovers the experimentally reported structure \cite{kyoto,kyoto_cif} after approximately 75 evaluations, while the remaining four plateau at slightly higher energies within the evaluation budget. This reflects the difficulty of the coupled search problem and highlights the need for multiple independent runs, given the stochastic nature of the search. The configuration space is also considerably larger than in the benchmark, comprising approximately $5\times10^{11}$ configurations before symmetry reduction, so recovering the experimental ordering within 75 evaluations demonstrates that the approach remains effective for multi-sublattice disorder. However, the tendency of most runs to did not reach the low-energy experimental ordering suggest that the current acquisition policy does not escape such basins reliably, and improving its exploration for coupled disorder is an important direction for future work.

\section{Conclusions}
In this work, we benchmarked three traditional global optimization methods (random walk, Markov chain Monte Carlo (MCMC), and a genetic algorithm (GA)) against four surrogate models (ExtraTrees ensemble, Bayesian ridge regression, sparse automatic relevance determination (ARD) regression, and an ensemble of ALiEGNN-inspired graph neural networks) within a shared agentic framework, with the aim of locating global-minimum structures of site-disordered crystals in as few evaluations as possible. Our results show that treating site-disorder resolution as an agentic global optimization problem, in which a surrogate fitted to the search's own evaluation history guides each subsequent relaxation, outperforms classical fixed-policy strategies. Across fourteen benchmark cases spanning twelve crystal structures, the agentic surrogates reach the enumerated global minimum in 94.3\% of runs, compared with 55.2\% for the traditional methods, and the overall best agentic surrogate, the Bayesian ridge regression, requires a median of 4.3-fold fewer evaluations than the best traditional method, the GA, outperforming it in every system. The choice of surrogate affects both reliability and efficiency. Sparse ARD regression is the fastest method in more than half of the systems but is less robust in others, whereas Bayesian ridge regression is the only method to succeed in every run and outperforms the GA in all fourteen systems, making it the recommended default. The framework also adapts to considerably more complex problems, as shown for \ce{Na_{2/3}FePO4}, where Na/vacancy configurational disorder and \ce{Fe^{2+}}/\ce{Fe^{3+}} charge disorder are coupled within a configuration space of approximately $5\times10^{11}$ configurations. Combining a charge-encoded surrogate representation with a charge-aware MLIP is sufficient for the Bayesian ridge search to recover the experimentally reported ordering, without prior knowledge of the target configuration or any change to the search policy. Together, these results show that, the efficiency of global optimization depends less on the cost of each evaluation than on how intelligently evaluations are chosen. Overall, our results demonstrate that agentic search substantially outperforms traditional optimization methods in both search efficiency and generalizability.


\section{Methods}
We benchmark seven global optimization strategies for recovering the global-minimum ordering of site-disordered crystal structures at fixed composition. As illustrated in \cref{fig:methods}, three traditional methods serve as non-adaptive baselines, while four agentic surrogate-guided procedures adapt online to their own observation history. All seven methods share the same energy evaluator, TensorNet-MATPES-PBE-v2\cite{Matgl,matpes}. Each candidate structure is relaxed with the LBFGS optimizer until the maximum residual force falls below \SI{0.02}{\electronvolt\per\angstrom}. Atomic positions and cell parameters are relaxed simultaneously using the FrechetCellFilter, with a symmetry constraint enforced throughout. All optimizations are performed with MatCalc\cite{Matgl} and the Atomic Simulation Environment (ASE)\cite{ASE}.

Prior to benchmarking, all orderings for each system were enumerated and reduced to the symmetry-inequivalent set using ASE's symmetry-equivalence tools (\cref{tab:systems}), establishing the global minimum as ground truth. Each method is allotted at most 200 energy evaluations per system, including initial structures (the initial structure and novelty-driven warm-up of the agentic methods, and the initial GA population). Previously evaluated occupancies are never re-evaluated; for computational efficiency, duplicates are identified by exact ordering, so the number of evaluations can exceed the number of symmetry-inequivalent configurations. A run is successful once it reaches the global-minimum structure or a symmetry-equivalent one with the same total energy. Each method is repeated over five independent runs.
\begin{table}[h]
\centering
\caption{Benchmark systems and their configuration spaces. For each system, the table gives the supercell, the number of disordered (swap) sites and their occupation, the total number of composition-preserving orderings, and the number of symmetry-inequivalent configurations remaining after symmetry reduction.}
\footnotesize
\label{tab:systems}
\begin{tabular}{@{}llcl rr@{}}
\toprule
System & Supercell & Swap sites & Occupation & Orderings & Inequivalent \\
\midrule
\ce{Al_{1.25}Ge_{0.75}Re}  & $2\times2\times1$ & 16 & Al 10, Ge 6  & 8\,008  & 122 \\
\ce{BaSc_{0.5}Ta_{0.5}O3}  & $2\times2\times4$ & 16 & Sc 8, Ta 8   & 12\,870 & 283 \\
\ce{CoPt}                  & $1\times2\times2$ & 16 & Co 8, Pt 8   & 12\,870 & 153 \\
\ce{Au_{0.8}Cu_{0.2}}      & $1\times1\times5$ & 20 & Au 16, Cu 4  & 4\,845  & 93  \\
\ce{Au_{0.8}Cu_{0.2}}      & $1\times2\times3$ & 24 & Au 19, Cu 5  & 42\,504 & 527 \\
\ce{FeCo}                  & $2\times2\times2$ & 16 & Co 8, Fe 8   & 12\,870 & 59  \\
\ce{FeCo}                  & $3\times3\times1$ & 18 & Co 9, Fe 9   & 48\,620 & 398 \\
\ce{Fe3Pt}                 & $1\times1\times5$ & 20 & Fe 15, Pt 5  & 15\,504 & 236 \\
\ce{CuInTe2}               & $1\times2\times2$ & 16 & Cu 8, In 8   & 12\,870 & 153 \\
\ce{LiYS2}                 & $1\times2\times2$ & 16 & Li 8, Y 8    & 12\,870 & 153 \\
\ce{Ni2SbTe2}              & $2\times2\times2$ & 16 & Sb 8, Te 8   & 12\,870 & 122 \\
\ce{ThCo3Mn2}              & $1\times1\times3$ & 15 & Co 9, Mn 6   & 5\,005  & 129 \\
\ce{TiCuNi}                & $1\times2\times2$ & 16 & Cu 8, Ni 8   & 12\,870 & 482 \\
\ce{Zn2VO4}                & $1\times1\times1$ & 16 & V 8, Zn 8    & 12\,870 & 97  \\
\bottomrule
\end{tabular}
\end{table}

\subsection{Traditional methods}
The random-walk baseline generates each new candidate by applying one to four randomly selected swaps between unlike occupants of the current occupancy. Energies are not used to direct proposals, except through a best-state restoration rule that returns the walk to the lowest-energy occupancy found so far when a proposed swap sequence returns to that occupancy from a different current occupancy, without re-evaluating it. 

The restricted MCMC baseline also proposes swap moves from the current occupancy, but accepts them according to
\begin{equation}
P_{\mathrm{acc}} =
\begin{cases}
1, & \Delta E \leq 0,\\
0.3, & \Delta E > 0,
\end{cases}
\end{equation}
where $\Delta E = E_{\mathrm{proposed}} - E_{\mathrm{current}}$. It differs from conventional Metropolis sampling in two respects: the uphill acceptance probability is fixed and independent of $\Delta E$, and previously evaluated occupancies are excluded from proposal, including those whose earlier moves were rejected. The method is therefore used here as an optimization baseline rather than a sampler of a stationary distribution.

The GA is initialized with a population of 20 occupancies generated by randomly distributing the disordered species over the available sites. In each generation, parents are chosen by roulette-wheel selection, favouring lower-energy (higher-fitness) structures. Offspring are generated by crossover, which combines the site occupancies of two parents, and by mutation, which permutes occupants among sites to maintain diversity. The GA runs for up to nine generations of up to 20 offspring each, targeting 200 relaxations in total. It terminates earlier if no novel offspring can be generated, and fewer relaxations may be completed if individual evaluations fail. No energy-based convergence criterion is used. Because candidates are evaluated in generations, the GA evaluation count is taken as the index of the first evaluation in the generation that contains the global minimum, which is a lower bound on the true count. Although these baselines retain information about the current state or population, none builds a model of the accumulated evaluation history to guide future proposals, in contrast to the agentic methods described below. The GA used in this work is build using the implementation in ASE-GA\cite{Lysgaard2013,Jennings2019,ASE}.

\subsection{Agentic surrogate methods}
Unlike the fixed-rule baselines, these methods select candidates using a stateful policy conditioned on the occupancies and relaxed energies observed in the current run. Here, \emph{agentic} refers to this adaptive numerical decision loop, where we utilize openai trough PUDA\cite{openai2026chatgpt,PUDA}.

\textit{Candidate generation.} The agent keeps an archive of up to four low-energy structures (elites), which must differ from each other at no fewer than four sites. New candidates are generated by swapping two unlike atoms in one of these elites, which preserves the composition. After eight proposals, the agent can also make larger jumps by swapping two, three, or four pairs of atoms at once. These jumps are made at least every five proposals, or earlier when all single-swap candidates have been evaluated. For each jump, up to 3000 unevaluated candidates are considered.

\textit{Descriptor and novelty.} For the tree and linear surrogates, an occupancy is represented by finite-template correlations inspired by cluster expansion and especially the clease package\cite{chang2019clease}. With $\sigma_i \in \{-1, +1\}$ encoding the species on substitutional site $i$, the descriptor comprises a constant term, the mean spin, pair correlations averaged within distance shells $\ell$, and triplet correlations averaged within triangle types $\tau$:
\begin{equation}
x(s) = \left[1,\ \overline{\sigma},\ \{\phi^{(2)}_{\ell}(s)\}_{\ell},\ \{\phi^{(3)}_{\tau}(s)\}_{\tau}\right]^{\mathsf{T}}
\end{equation}
The novelty $\nu(s)$ of a candidate is the Euclidean distance to the nearest evaluated descriptor after feature scaling, with scale factors computed from the combined history and candidate pool (floored at 0.25). The graph surrogate predicts energies from a fixed site graph but uses the same descriptor-based novelty for selection.

\textit{Acquisition.} While fewer than eight energies are available, the candidate with maximum novelty is selected. Thereafter, candidates are selected by minimizing the lower-confidence-bound score
\begin{equation}
a_t(s) = \mu_t(s) - 1.5\,u_t(s) - 0.08\,s_E\,\tilde{\nu}(s)
\end{equation}
where $s_E = \max\{\mathrm{std}(E_1, \ldots, E_t),\ 10^{-3}\,\mathrm{eV}\}$ and $\tilde{\nu}$ is the novelty normalized by its maximum over the current pool. For single-exchange pools, uncertainty selection ($a_t = -u_t$) and greedy selection ($a_t = \mu_t$) replace this score when the step count is divisible by 17 and 29, respectively. Coordinated exchanges rotate through the lower-confidence-bound, greedy, uncertainty, and novelty modes, and these required modes take precedence. Every evaluated candidate is added to the run history, whether or not it improves the best energy.

\textit{Stopping.} The initial occupancy is evaluated at step 0, after which a run makes at most 200 proposals. It may stop earlier, after at least 32 proposals, only if all of the following hold: all single-exchange neighbourhoods of the current elites have been evaluated; every required acquisition mode for each coordinated-exchange size and elite has been used, or that jump space is exhausted; the best energy has not improved for 40 proposals; and the elite archive has been unchanged for 25 proposals. Exhaustion of the candidate space also ends a run. Neither condition guarantees that the global minimum has been found.

\textit{Surrogate models.} The first three surrogates use scikit-learn\cite{pedregosa2011scikit}. The linear models centre and scale features using the evaluated history only, whereas ExtraTrees uses the candidate-pool scaling described above. The fourth is a compact graph ensemble inspired by ALiEGNN\cite{klat_surrogate}, not the full pretrained architecture:
\begin{itemize}
\item \textbf{ExtraTrees}: 400 extremely randomized trees (minimum leaf size 2, feature fraction 0.9). $\mu_t$ and $u_t$ are the mean and standard deviation of the tree predictions; $u_t$ is an uncalibrated measure of tree disagreement.
\item \textbf{Bayesian ridge regression}: a linear model with an isotropic Gaussian coefficient prior and evidence-estimated hyperparameters; $\mu_t$ and $u_t$ are the predictive mean and standard deviation.
\item \textbf{Sparse ARD regression}: a linear model with a separate prior precision for each coefficient, using a pruning threshold of $10^4$.
\item \textbf{ALiEGNN-inspired ensemble}: three compact networks (3153 trainable parameters each), trained from scratch on bootstrap samples of the run history at each model-based selection (20 AdamW epochs). A fixed graph connects each substitutional site to at most 12 periodic neighbours within 8~\AA, and its line graph encodes bond-angle cosines. Each network uses species embeddings, radial bond features, one angular line-graph and gated atom message-passing block, mean pooling, and a scalar energy head, without pretrained weights or E3NN coordinate updates. $\mu_t$ is the ensemble mean, and $u_t$ combines the ensemble variance with the in-sample root-mean-square residual as an uncalibrated uncertainty proxy.
\end{itemize}

\subsection{\ce{Na_{2/3}FePO4} case study}
Following Ref.~\cite{charge_ordering}, the orthorhombic olivine \ce{NaFePO4} cell was expanded to a $1\times3\times2$ supercell containing 24 Na and 24 Fe sites. At \ce{Na_{2/3}FePO4}, 16 Na sites are occupied and 8 are vacant, and charge neutrality requires 8 of the 24 Fe ions to be \ce{Fe^{3+}} and the remaining 16 to be \ce{Fe^{2+}}. The joint configuration space therefore comprises $\binom{24}{8}^2 \approx 5.4\times10^{11}$ configurations before symmetry reduction, far beyond exhaustive enumeration. Energies were evaluated with the charge-state-aware MACE model of Ref.~\cite{charge_ordering}, which assigns separate encoding to \ce{Fe^{2+}} and \ce{Fe^{3+}} and was trained on PBE+$U$ ($U_{\mathrm{Fe}} = 5.3$~eV) Fe-based dataset. In the surrogate representation, \ce{Fe^{2+}} and \ce{Fe^{3+}} are likewise treated as distinct species, so that the search operates jointly over the Na/vacancy and Fe charge sublattices. The experimentally reported ordering \cite{kyoto,kyoto_cif}, originally monoclinic, was mapped onto the orthorhombic lattice for comparison, as in Ref.~\cite{charge_ordering}. The Bayesian ridge surrogate and acquisition settings were identical to those of the benchmark, with five independent runs of up to 200 proposals each.

\textbf{Acknowledgements}
This research is supported by the National Research Foundation, Prime Minister’s Office, Singapore under its Campus for Research Excellence and Technological Enterprise (CREATE) program. M. H. P. acknowledges the National Research Foundation, Singapore and their AI-for-x postdoctoral fellowship initiative for their financial support. 

\textbf{Data and code accessibility}
The code used for this work is available at \url{https://github.com/mahpe/Agenticoptimizer} and the enumerated symmetrical in-equivalent structural optimized systems are available at \url{https://huggingface.co/datasets/Mahpe/Agentic_Global_optimization}.

\bibliography{refs}
\newpage

\begin{appendices}

\section{Convergence trajectories for all benchmark systems}
\label{app:all_runs}
\begin{figure}[b!]
    \centering
    \includegraphics[width=1\textwidth]{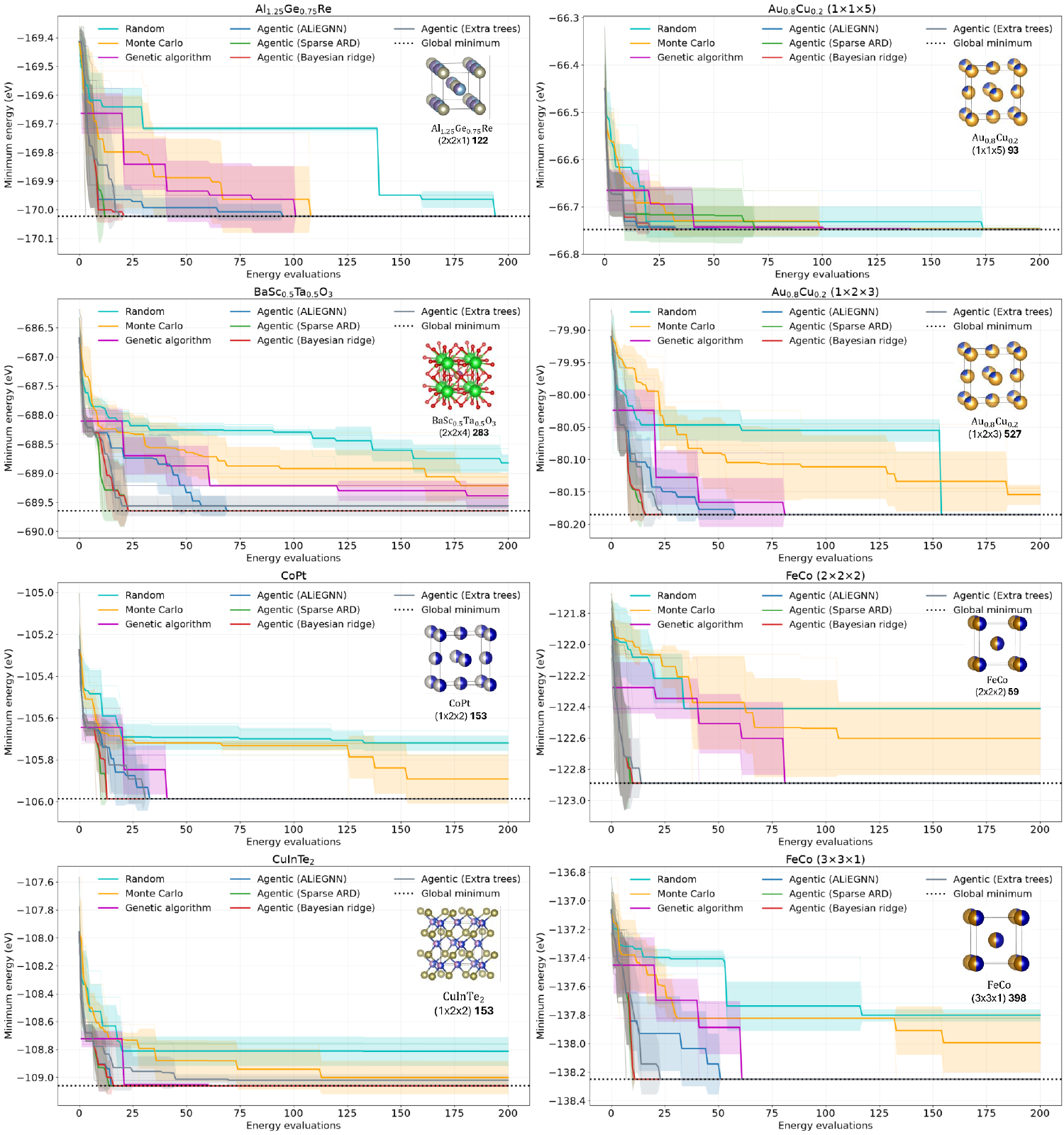}
    \caption{Best-so-far minimum energy versus number of energy evaluations for all seven methods on \ce{Al_{1.25}Ge_{0.75}Re}, \ce{Au_{0.8}Cu_{0.2}} ($1\times1\times5$ and $1\times2\times3$), \ce{BaSc_{0.5}Ta_{0.5}O3}, \ce{CoPt}, \ce{FeCo} ($2\times2\times2$ and $3\times3\times1$), and \ce{CuInTe2}. Curves show the mean over five independent runs, with $\pm1\sigma$ shaded bands. The dotted line marks the enumerated global minimum. Insets show the structure, supercell, and number of symmetry-inequivalent configurations.}
    \label{fig:all_runs1}
\end{figure}

\begin{figure}[b!]
    \centering
    \includegraphics[width=1\textwidth]{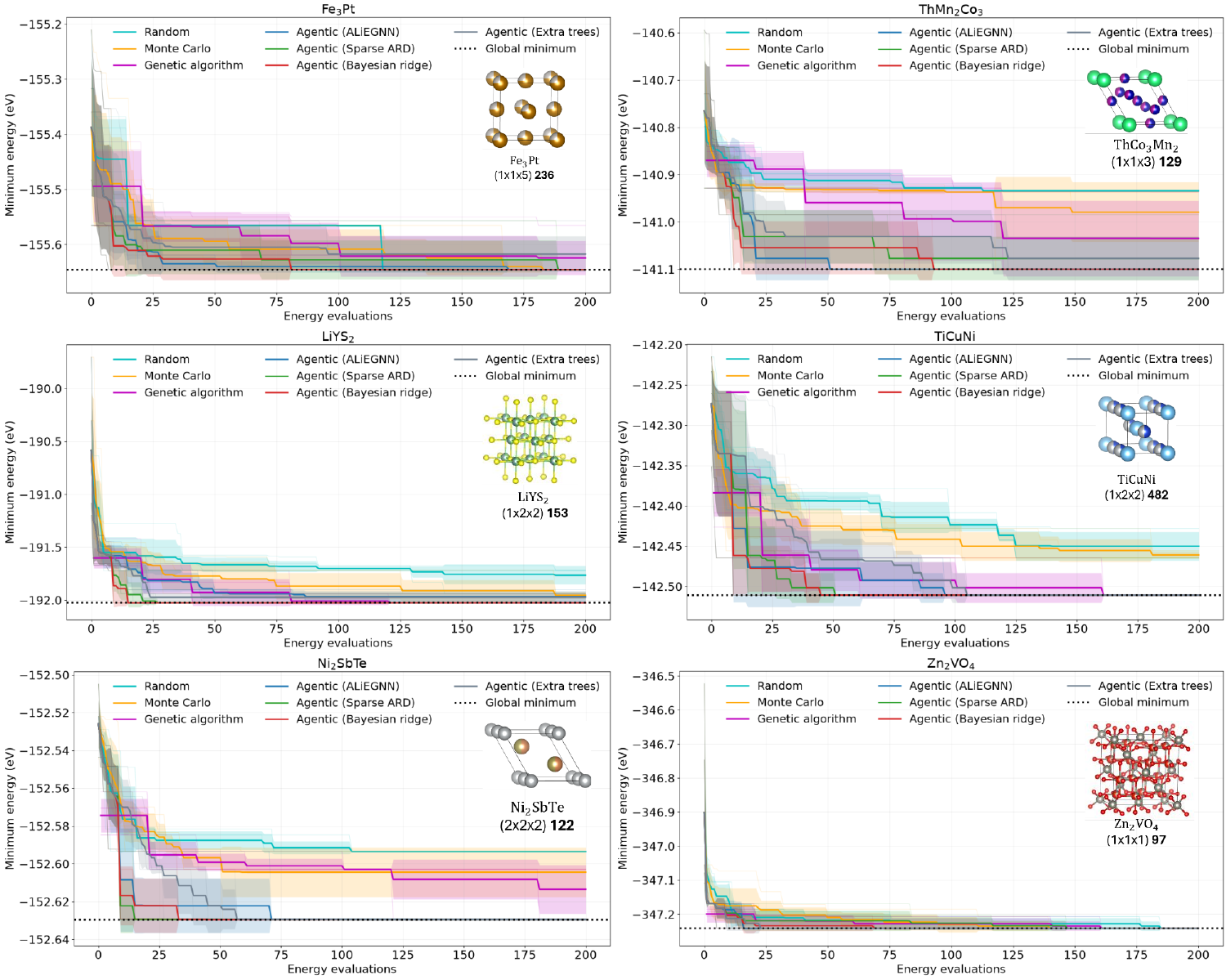}
    \caption{As in \cref{fig:all_runs1}, for \ce{Fe3Pt}, \ce{ThCo3Mn2}, \ce{LiYS2}, \ce{TiCuNi}, \ce{Ni2SbTe2}, and \ce{Zn2VO4}.}
    \label{fig:all_runs2}
\end{figure}

\Cref{fig:all_runs1,fig:all_runs2} show the best-so-far minimum energy as a function of the number of energy evaluations for all seven methods on each of the fourteen benchmark cases. Each curve is the mean over five independent runs with the shaded bands indicating $\pm1\sigma$, and the dotted line the enumerated global minimum. The inset in each panel shows the crystal structure, the supercell, and the number of symmetry-inequivalent configurations. \Cref{fig:example} in the main text corresponds to the \ce{Au_{0.8}Cu_{0.2}} ($1\times2\times3$) panel.

The trajectories reproduce the trends summarized in \cref{fig:histogram} and \cref{tab:stats}. In most systems, the agentic surrogates drop to the global minimum within the first few tens of evaluations, typically shortly after the eight-structure warm-up, whereas random walk and MCMC frequently plateau above the global minimum for the remainder of the budget. This separation is most pronounced for systems with a large energy spread between configurations, such as \ce{BaSc_{0.5}Ta_{0.5}O3}, \ce{CoPt}, \ce{CuInTe2}, both \ce{FeCo} supercells, \ce{TiCuNi}, \ce{ThCo3Mn2}, and \ce{Ni2SbTe2}. The GA typically descends in discrete steps between generations, giving the characteristic plateaus discussed for \ce{Au_{0.8}Cu_{0.2}}. For systems with a narrow energy distribution, such as \ce{LiYS2}, \ce{Zn2VO4}, and \ce{Au_{0.8}Cu_{0.2}} ($1\times1\times5$), all methods quickly approach the global minimum in energy, and differences between methods are mainly seen in whether the exact global-minimum structure is located. The larger spread across runs for ExtraTrees and the ALiEGNN-inspired ensemble in some systems, including \ce{Ni2SbTe2} and \ce{ThCo3Mn2}, reflects the greater seed-to-seed variability of these surrogates.

\section{Per-run results for all methods}
\label{app:seeds}
\Cref{tab:seeds_rw,tab:seeds_mcmc,tab:seeds_ga,tab:seeds_br,tab:seeds_ard,tab:seeds_aliegnn,tab:seeds_et} list, for each method, the number of energy evaluations required by each of the five independent runs to first reach the enumerated global minimum in every benchmark system. These tables show the per-system means and success rates summarized in \cref{tab:stats}. For the GA, which evaluates candidates in generations of 20, the reported value is the first evaluation of the generation in which the global minimum was first reached and is therefore a lower bound on the true evaluation count.

The per-run results highlight the difference in reliability between the two method classes. The traditional baselines frequently fail in several of the five runs, and random walk succeeds in only five of the fourteen systems. Bayesian ridge succeeds in every run, with most runs requiring fewer than 20 evaluations; its slowest runs (82 evaluations on \ce{Fe3Pt}, and 88 and 94 on \ce{ThCo3Mn2}) remain well within the budget. Sparse ARD shows a similar distribution but occasionally requires considerably more evaluations, for example 190 evaluations in one \ce{Fe3Pt} run and 148 in one \ce{Zn2VO4} run, reflecting the larger run-to-run variability. ExtraTrees and the ALiEGNN-inspired ensemble also succeed in most runs, but fail in up to three of five runs on \ce{Fe3Pt}, \ce{CuInTe2}, and \ce{LiYS2}.

\begin{table}[h!]
\centering
\caption{Number of energy evaluations required by each of the five independent runs of \textbf{the random-walk} baseline to first reach the enumerated global minimum. A dash indicates that the run did not reach the global minimum within the evaluation budget. The last two columns give the number of successful runs and the mean $\pm$ standard deviation over successful runs.}
\label{tab:seeds_rw}
\footnotesize
\begin{tabular}{@{}lccccc|cc@{}}
\toprule
System & Run 1 & Run 2 & Run 3 & Run 4 & Run 5 & Success & Mean $\pm$ std \\
\midrule
\ce{Al_{1.25}Ge_{0.75}Re} & 195 & 195 & 195 & 161 & 195 & 5/5 & 188.2 $\pm$ 13.6 \\
\ce{BaSc_{0.5}Ta_{0.5}O3} & --- & --- & --- & --- & --- & 0/5 & --- \\
\ce{CoPt} & --- & --- & --- & --- & --- & 0/5 & --- \\
\ce{Au_{0.8}Cu_{0.2}} ($1{\times}1{\times}5$) & 20 & 20 & 20 & --- & 20 & 4/5 & 20.0 $\pm$ 0.0 \\
\ce{Au_{0.8}Cu_{0.2}} ($1{\times}2{\times}3$) & 155 & 155 & 155 & 155 & 155 & 5/5 & 155.0 $\pm$ 0.0 \\
\ce{FeCo} ($2{\times}2{\times}2$) & --- & --- & --- & --- & --- & 0/5 & --- \\
\ce{FeCo} ($3{\times}3{\times}1$) & --- & --- & --- & --- & --- & 0/5 & --- \\
\ce{Fe3Pt} & 119 & 119 & 119 & 119 & 119 & 5/5 & 119.0 $\pm$ 0.0 \\
\ce{CuInTe2} & --- & --- & --- & --- & --- & 0/5 & --- \\
\ce{LiYS2} & --- & --- & --- & --- & --- & 0/5 & --- \\
\ce{Ni2SbTe2} & --- & --- & --- & --- & --- & 0/5 & --- \\
\ce{ThCo3Mn2} & --- & --- & --- & --- & --- & 0/5 & --- \\
\ce{TiCuNi} & --- & --- & --- & --- & --- & 0/5 & --- \\
\ce{Zn2VO4} & 12 & 11 & 186 & --- & 178 & 4/5 & 96.8 $\pm$ 85.3 \\
\bottomrule
\end{tabular}
\end{table}

\begin{table}[h]
\centering
\caption{Number of energy evaluations required by each of the five independent runs of \textbf{the Markov chain Monte Carlo} baseline to first reach the enumerated global minimum. A dash indicates that the run did not reach the global minimum within the evaluation budget. The last two columns give the number of successful runs and the mean $\pm$ standard deviation over successful runs.}
\label{tab:seeds_mcmc}
\footnotesize
\begin{tabular}{@{}lccccc|cc@{}}
\toprule
System & Run 1 & Run 2 & Run 3 & Run 4 & Run 5 & Success & Mean $\pm$ std \\
\midrule
\ce{Al_{1.25}Ge_{0.75}Re} & 68 & 63 & 10 & 109 & 8 & 5/5 & 51.6 $\pm$ 38.3 \\
\ce{BaSc_{0.5}Ta_{0.5}O3} & --- & --- & --- & 163 & --- & 1/5 & 163.0 $\pm$ 0.0 \\
\ce{CoPt} & 154 & --- & 139 & 127 & --- & 3/5 & 140.0 $\pm$ 11.0 \\
\ce{Au_{0.8}Cu_{0.2}} ($1{\times}1{\times}5$) & --- & 33 & 100 & 15 & 15 & 4/5 & 40.8 $\pm$ 35.0 \\
\ce{Au_{0.8}Cu_{0.2}} ($1{\times}2{\times}3$) & --- & --- & --- & --- & 44 & 1/5 & 44.0 $\pm$ 0.0 \\
\ce{FeCo} ($2{\times}2{\times}2$) & 68 & --- & --- & 39 & --- & 2/5 & 53.5 $\pm$ 14.5 \\
\ce{FeCo} ($3{\times}3{\times}1$) & 156 & --- & 134 & --- & --- & 2/5 & 145.0 $\pm$ 11.0 \\
\ce{Fe3Pt} & 47 & 168 & 184 & 19 & 137 & 5/5 & 111.0 $\pm$ 66.1 \\
\ce{CuInTe2} & 114 & 75 & --- & 37 & 29 & 4/5 & 63.8 $\pm$ 33.8 \\
\ce{LiYS2} & 54 & --- & --- & --- & --- & 1/5 & 54.0 $\pm$ 0.0 \\
\ce{Ni2SbTe2} & --- & --- & --- & --- & --- & 0/5 & --- \\
\ce{ThCo3Mn2} & --- & --- & --- & --- & 119 & 1/5 & 119.0 $\pm$ 0.0 \\
\ce{TiCuNi} & --- & --- & --- & --- & --- & 0/5 & --- \\
\ce{Zn2VO4} & 57 & 32 & 96 & 141 & 74 & 5/5 & 80.0 $\pm$ 37.0 \\
\bottomrule
\end{tabular}
\end{table}

\begin{table}[h]
\centering
\caption{Number of energy evaluations required by each of the five independent runs of \textbf{the genetic algorithm (GA)} to first reach the enumerated global minimum. A dash indicates that the run did not reach the global minimum within the evaluation budget. The last two columns give the number of successful runs and the mean $\pm$ standard deviation over successful runs. For the GA, the value is the first evaluation of the generation in which the global minimum was first reached.}
\label{tab:seeds_ga}
\footnotesize
\begin{tabular}{@{}lccccc|cc@{}}
\toprule
System & Run 1 & Run 2 & Run 3 & Run 4 & Run 5 & Success & Mean $\pm$ std \\
\midrule
\ce{Al_{1.25}Ge_{0.75}Re} & 41 & 81 & 41 & 61 & 101 & 5/5 & 65.0 $\pm$ 23.3 \\
\ce{BaSc_{0.5}Ta_{0.5}O3} & --- & --- & 181 & 121 & --- & 2/5 & 151.0 $\pm$ 30.0 \\
\ce{CoPt} & 41 & 41 & 21 & 21 & 41 & 5/5 & 33.0 $\pm$ 9.8 \\
\ce{Au_{0.8}Cu_{0.2}} ($1{\times}1{\times}5$) & 101 & 101 & 141 & 41 & 41 & 5/5 & 85.0 $\pm$ 38.8 \\
\ce{Au_{0.8}Cu_{0.2}} ($1{\times}2{\times}3$) & 21 & 41 & 41 & 81 & 41 & 5/5 & 45.0 $\pm$ 19.6 \\
\ce{FeCo} ($2{\times}2{\times}2$) & 61 & 81 & 41 & 81 & 81 & 5/5 & 69.0 $\pm$ 16.0 \\
\ce{FeCo} ($3{\times}3{\times}1$) & 61 & 61 & 61 & 61 & 41 & 5/5 & 57.0 $\pm$ 8.0 \\
\ce{Fe3Pt} & 61 & --- & 101 & 101 & --- & 3/5 & 87.7 $\pm$ 18.9 \\
\ce{CuInTe2} & 21 & 21 & 61 & 21 & 21 & 5/5 & 29.0 $\pm$ 16.0 \\
\ce{LiYS2} & 41 & 41 & 21 & 81 & 121 & 5/5 & 61.0 $\pm$ 35.8 \\
\ce{Ni2SbTe2} & 181 & --- & --- & --- & --- & 1/5 & 181.0 $\pm$ 0.0 \\
\ce{ThCo3Mn2} & 81 & 121 & 41 & --- & --- & 3/5 & 81.0 $\pm$ 32.7 \\
\ce{TiCuNi} & 161 & 21 & 101 & 61 & 21 & 5/5 & 73.0 $\pm$ 53.1 \\
\ce{Zn2VO4} & 1 & 161 & 21 & 141 & 61 & 5/5 & 77.0 $\pm$ 63.7 \\
\bottomrule
\end{tabular}
\end{table}

\begin{table}[h]
\centering
\caption{Number of energy evaluations required by each of the five independent runs of \textbf{the Bayesian ridge surrogate} to first reach the enumerated global minimum. A dash indicates that the run did not reach the global minimum within the evaluation budget. The last two columns give the number of successful runs and the mean $\pm$ standard deviation over successful runs.}
\label{tab:seeds_br}
\footnotesize
\begin{tabular}{@{}lccccc|cc@{}}
\toprule
System & Run 1 & Run 2 & Run 3 & Run 4 & Run 5 & Success & Mean $\pm$ std \\
\midrule
\ce{Al_{1.25}Ge_{0.75}Re} & 22 & 10 & 10 & 4 & 10 & 5/5 & 11.2 $\pm$ 5.9 \\
\ce{BaSc_{0.5}Ta_{0.5}O3} & 12 & 17 & 14 & 24 & 20 & 5/5 & 17.4 $\pm$ 4.3 \\
\ce{CoPt} & 14 & 14 & 11 & 9 & 14 & 5/5 & 12.4 $\pm$ 2.1 \\
\ce{Au_{0.8}Cu_{0.2}} ($1{\times}1{\times}5$) & 3 & 16 & 22 & 10 & 22 & 5/5 & 14.6 $\pm$ 7.3 \\
\ce{Au_{0.8}Cu_{0.2}} ($1{\times}2{\times}3$) & 16 & 17 & 9 & 9 & 10 & 5/5 & 12.2 $\pm$ 3.5 \\
\ce{FeCo} ($2{\times}2{\times}2$) & 3 & 11 & 7 & 3 & 5 & 5/5 & 5.8 $\pm$ 3.0 \\
\ce{FeCo} ($3{\times}3{\times}1$) & 8 & 11 & 12 & 11 & 11 & 5/5 & 10.6 $\pm$ 1.4 \\
\ce{Fe3Pt} & 8 & 9 & 82 & 10 & 30 & 5/5 & 27.8 $\pm$ 28.3 \\
\ce{CuInTe2} & 14 & 17 & 10 & 10 & 10 & 5/5 & 12.2 $\pm$ 2.9 \\
\ce{LiYS2} & 16 & 10 & 10 & 16 & 10 & 5/5 & 12.4 $\pm$ 2.9 \\
\ce{Ni2SbTe2} & 34 & 46 & 34 & 46 & 28 & 5/5 & 37.6 $\pm$ 7.2 \\
\ce{ThCo3Mn2} & 13 & 88 & 94 & 16 & 13 & 5/5 & 44.8 $\pm$ 37.8 \\
\ce{TiCuNi} & 10 & 10 & 40 & 40 & 46 & 5/5 & 29.2 $\pm$ 15.8 \\
\ce{Zn2VO4} & 70 & 15 & 17 & 17 & 6 & 5/5 & 25.0 $\pm$ 22.9 \\
\bottomrule
\end{tabular}
\end{table}

\begin{table}[h]
\centering
\caption{Number of energy evaluations required by each of the five independent runs of \textbf{the sparse ARD surrogate} to first reach the enumerated global minimum. A dash indicates that the run did not reach the global minimum within the evaluation budget. The last two columns give the number of successful runs and the mean $\pm$ standard deviation over successful runs.}
\label{tab:seeds_ard}
\footnotesize
\begin{tabular}{@{}lccccc|cc@{}}
\toprule
System & Run 1 & Run 2 & Run 3 & Run 4 & Run 5 & Success & Mean $\pm$ std \\
\midrule
\ce{Al_{1.25}Ge_{0.75}Re} & 10 & 10 & 10 & 4 & 13 & 5/5 & 9.4 $\pm$ 2.9 \\
\ce{BaSc_{0.5}Ta_{0.5}O3} & 11 & 11 & 24 & 12 & 20 & 5/5 & 15.6 $\pm$ 5.4 \\
\ce{CoPt} & 14 & 11 & 11 & 9 & 14 & 5/5 & 11.8 $\pm$ 1.9 \\
\ce{Au_{0.8}Cu_{0.2}} ($1{\times}1{\times}5$) & 3 & 22 & 88 & 10 & 52 & 5/5 & 35.0 $\pm$ 31.4 \\
\ce{Au_{0.8}Cu_{0.2}} ($1{\times}2{\times}3$) & 17 & 14 & 9 & 9 & 10 & 5/5 & 11.8 $\pm$ 3.2 \\
\ce{FeCo} ($2{\times}2{\times}2$) & 3 & 10 & 7 & 3 & 5 & 5/5 & 5.6 $\pm$ 2.7 \\
\ce{FeCo} ($3{\times}3{\times}1$) & 8 & 11 & 12 & 11 & 10 & 5/5 & 10.4 $\pm$ 1.4 \\
\ce{Fe3Pt} & 8 & 9 & 70 & 190 & 15 & 5/5 & 58.4 $\pm$ 69.7 \\
\ce{CuInTe2} & 14 & 15 & 13 & 10 & 10 & 5/5 & 12.4 $\pm$ 2.1 \\
\ce{LiYS2} & 28 & 22 & 10 & 16 & 16 & 5/5 & 18.4 $\pm$ 6.1 \\
\ce{Ni2SbTe2} & 10 & 70 & 52 & 10 & 28 & 5/5 & 34.0 $\pm$ 23.7 \\
\ce{ThCo3Mn2} & --- & 76 & 70 & 17 & 13 & 4/5 & 44.0 $\pm$ 29.1 \\
\ce{TiCuNi} & 28 & 10 & 52 & 16 & 34 & 5/5 & 28.0 $\pm$ 14.7 \\
\ce{Zn2VO4} & 148 & 15 & 17 & 118 & 6 & 5/5 & 60.8 $\pm$ 59.8 \\
\bottomrule
\end{tabular}
\end{table}

\begin{table}[h]
\centering
\caption{Number of energy evaluations required by each of the five independent runs of \textbf{the ALiEGNN-inspired ensemble} to first reach the enumerated global minimum. A dash indicates that the run did not reach the global minimum within the evaluation budget. The last two columns give the number of successful runs and the mean $\pm$ standard deviation over successful runs.}
\label{tab:seeds_aliegnn}
\footnotesize
\begin{tabular}{@{}lccccc|cc@{}}
\toprule
System & Run 1 & Run 2 & Run 3 & Run 4 & Run 5 & Success & Mean $\pm$ std \\
\midrule
\ce{Al_{1.25}Ge_{0.75}Re} & 66 & 96 & 31 & 4 & 10 & 5/5 & 41.4 $\pm$ 34.9 \\
\ce{BaSc_{0.5}Ta_{0.5}O3} & 38 & 54 & 45 & 58 & 70 & 5/5 & 53.0 $\pm$ 11.0 \\
\ce{CoPt} & 28 & 12 & 18 & 16 & 34 & 5/5 & 21.6 $\pm$ 8.1 \\
\ce{Au_{0.8}Cu_{0.2}} ($1{\times}1{\times}5$) & 3 & 54 & 16 & 34 & 34 & 5/5 & 28.2 $\pm$ 17.4 \\
\ce{Au_{0.8}Cu_{0.2}} ($1{\times}2{\times}3$) & 25 & 59 & 18 & 32 & 42 & 5/5 & 35.2 $\pm$ 14.3 \\
\ce{FeCo} ($2{\times}2{\times}2$) & 3 & 11 & 7 & 3 & 5 & 5/5 & 5.8 $\pm$ 3.0 \\
\ce{FeCo} ($3{\times}3{\times}1$) & 8 & 46 & 34 & 52 & 14 & 5/5 & 30.8 $\pm$ 17.3 \\
\ce{Fe3Pt} & 8 & 170 & 52 & 20 & 16 & 5/5 & 53.2 $\pm$ 60.3 \\
\ce{CuInTe2} & 16 & 10 & 9 & 10 & 16 & 5/5 & 12.2 $\pm$ 3.1 \\
\ce{LiYS2} & 46 & --- & --- & 88 & --- & 2/5 & 67.0 $\pm$ 21.0 \\
\ce{Ni2SbTe2} & 10 & --- & 188 & 129 & 10 & 4/5 & 84.2 $\pm$ 77.1 \\
\ce{ThCo3Mn2} & 17 & 22 & 22 & 52 & 12 & 5/5 & 25.0 $\pm$ 14.0 \\
\ce{TiCuNi} & 10 & 10 & 87 & 16 & 97 & 5/5 & 44.0 $\pm$ 39.4 \\
\ce{Zn2VO4} & 17 & 17 & 17 & 14 & 6 & 5/5 & 14.2 $\pm$ 4.3 \\
\bottomrule
\end{tabular}
\end{table}

\begin{table}[h]
\centering
\caption{Number of energy evaluations required by each of the five independent runs of \textbf{the ExtraTrees surrogate} to first reach the enumerated global minimum. A dash indicates that the run did not reach the global minimum within the evaluation budget. The last two columns give the number of successful runs and the mean $\pm$ standard deviation over successful runs.}
\label{tab:seeds_et}
\footnotesize
\begin{tabular}{@{}lccccc|cc@{}}
\toprule
System & Run 1 & Run 2 & Run 3 & Run 4 & Run 5 & Success & Mean $\pm$ std \\
\midrule
\ce{Al_{1.25}Ge_{0.75}Re} & 10 & 18 & 20 & 4 & 18 & 5/5 & 14.0 $\pm$ 6.1 \\
\ce{BaSc_{0.5}Ta_{0.5}O3} & --- & 21 & 13 & 18 & 17 & 4/5 & 17.2 $\pm$ 2.9 \\
\ce{CoPt} & 32 & 11 & 24 & 15 & 31 & 5/5 & 22.6 $\pm$ 8.4 \\
\ce{Au_{0.8}Cu_{0.2}} ($1{\times}1{\times}5$) & 3 & 40 & 22 & 16 & 34 & 5/5 & 23.0 $\pm$ 13.1 \\
\ce{Au_{0.8}Cu_{0.2}} ($1{\times}2{\times}3$) & 16 & 23 & 9 & 25 & 16 & 5/5 & 17.8 $\pm$ 5.7 \\
\ce{FeCo} ($2{\times}2{\times}2$) & 3 & 15 & 7 & 3 & 5 & 5/5 & 6.6 $\pm$ 4.5 \\
\ce{FeCo} ($3{\times}3{\times}1$) & 8 & 12 & 24 & 15 & 15 & 5/5 & 14.8 $\pm$ 5.3 \\
\ce{Fe3Pt} & 8 & 22 & --- & --- & --- & 2/5 & 15.0 $\pm$ 7.0 \\
\ce{CuInTe2} & --- & 71 & 40 & --- & --- & 2/5 & 55.5 $\pm$ 15.5 \\
\ce{LiYS2} & --- & 18 & --- & --- & 22 & 2/5 & 20.0 $\pm$ 2.0 \\
\ce{Ni2SbTe2} & 64 & 112 & 100 & 70 & 70 & 5/5 & 83.2 $\pm$ 19.1 \\
\ce{ThCo3Mn2} & 124 & 26 & 118 & 17 & --- & 4/5 & 71.2 $\pm$ 49.9 \\
\ce{TiCuNi} & 88 & 106 & 84 & 100 & 46 & 5/5 & 84.8 $\pm$ 21.0 \\
\ce{Zn2VO4} & 18 & 24 & 10 & 10 & 6 & 5/5 & 13.6 $\pm$ 6.5 \\
\bottomrule
\end{tabular}
\end{table}

\end{appendices}

\end{document}